\documentclass[lettersize,journal]{IEEEtran}
\usepackage{amsmath,amsfonts}
\usepackage{algorithmic}
\usepackage{algorithm}
\usepackage{array}
\usepackage[caption=false,font=normalsize,labelfont=sf,textfont=sf]{subfig}
\usepackage{textcomp}
\usepackage{stfloats}
\usepackage{url}
\usepackage{verbatim}
\usepackage{graphicx}
\usepackage{cite}

\usepackage[utf8]{inputenc}
\usepackage{booktabs}
\usepackage{multirow} 

\usepackage{amsmath}
\usepackage{amssymb} 
\usepackage{bm}      

\DeclareMathOperator*{\argmin}{argmin}

\begin{document}

\title{Propagation-Consistent Wireless Environment Digital Twin Construction Under Sparse Measurements}

\author{Junjie Ai,~\IEEEmembership{Graduate Student Member,~IEEE}, Shurui Xu, Yanqing Ren, Zhuoyu Liu, Weicong Chen,~\IEEEmembership{Member,~IEEE}, Wankai Tang,~\IEEEmembership{Member,~IEEE}, Xiao Li,~\IEEEmembership{Member,~IEEE}, Chao-Kai Wen,~\IEEEmembership{Fellow,~IEEE}, and Shi Jin,~\IEEEmembership{Fellow,~IEEE}
\thanks{\textit{(Corresponding authors: Shi Jin; Weicong Chen).}}
\thanks{J. Ai, S. Xu, Y. Ren, Z. Liu, W. Chen, W. Tang, X. Li, and S. Jin are with the School of Information Science and Engineering, Southeast University, Nanjing 210096, China (e-mail: aijj@seu.edu.cn; 213221669@seu.edu.cn; yq\_ren@seu.edu.cn; 213242912@seu.edu.cn; cwc@seu.edu.cn; tangwk@seu.edu.cn; li\_xiao@seu.edu.cn; jinshi@seu.edu.cn).}
\thanks{C.-K. Wen is with the Institute of Communications Engineering, National Sun Yat-sen University, Kaohsiung 80424, Taiwan (e-mail: chaokai.wen@mail.nsysu.edu.tw).}
}


\markboth{Journal of \LaTeX\ Class Files,~Vol.~14, No.~8, August~2021}%
{Shell \MakeLowercase{\textit{et al.}}: A Sample Article Using IEEEtran.cls for IEEE Journals}


\maketitle
\vspace{-2cm}
\begin{abstract}

Digital twins (DTs) are promising for wireless deployment, optimization, and data generation, but building a propagation-faithful twin from sparse real measurements remains difficult. 
This paper proposes a wireless environment digital twin (WEDT) construction paradigm that evolves a reconstructed geometric DT into a propagation-consistent wireless environment representation through calibration of a scene-level electromagnetic (EM) property field. 
Instead of directly fitting link-specific channel responses, the proposed paradigm first constructs a geometry-prior Bayesian channel map (BCM) to convert sparse position-labeled channel state information (CSI) into dense probabilistic supervision with uncertainty estimates. 
It then embeds the learnable EM property field into differentiable ray tracing (RT) based channel computation, thereby enabling calibration through an explicit propagation chain. 
Experiments in both public and real-world scenes show that WEDT achieves accurate channel prediction, generalizes to unseen transceiver topologies, and remains effective across different sampling conditions. 
WEDT also offers utility for material-related environment sensing, more reliable physical-layer planning, and higher-quality synthetic data generation for wireless AI. 
These results demonstrate the value of the proposed paradigm for propagation-consistent WEDT construction and related wireless applications.

\end{abstract}


\section{Introduction}
\IEEEPARstart{A}{s} wireless devices proliferate and communication environments become increasingly complex, sixth-generation (6G) wireless systems are expected to face growing demands on spectral efficiency, wide-area coverage, and link reliability \cite{6G}.
To meet these demands, the research community has explored a series of emerging 6G technologies, such as reconfigurable intelligent surfaces (RIS), integrated sensing and communication (ISAC), and artificial intelligence (AI) native wireless systems \cite{RIS,ISAC,AI-comm}, driving the evolution of wireless systems towards more adaptive and intelligent architectures.
However, the practical implementation of these technologies in complex real-world environments remains highly challenging.
On the one hand, wireless system deployment and optimization in real-world environments still rely heavily on expert knowledge, field testing, and repeated trial-and-error, leading to long deployment cycles and high operational costs \cite{6G_deployment_challenge}.
On the other hand, AI-driven wireless methods require large amounts of high-quality, well-labeled channel data for model training and evaluation, yet such data are costly and time-consuming to collect in practice \cite{6G_data_challenge}.
As a result, future wireless systems face a dual bottleneck: costly real-world deployment and optimization, and scarce channel data.  

As digital representations of real world, Digital twin (DTs) offers a highly promising solution to address the aforementioned bottlenecks in wireless communications \cite{DT-comm}.
By constructing simulation platforms of real-world environments, DT enables the online configuration, analysis, and optimization of wireless systems, thereby reducing reliance on costly field experiments and accelerating their practical deployment \cite{DT_deployment}.
Furthermore, it can generate vast quantities of labeled synthetic data for specific environments, effectively alleviating the data scarcity issues faced in the training and evaluation of AI-based wireless communication methods \cite{DT_data}.
To fully realize the application potential of DT, however, the key challenge is how to construct appropriate and physically faithful representations of real-world wireless communications.

In recent years, driven by the development of mobile terminal sensor technologies and the advancement of three-dimensional (3D) reconstruction algorithms \cite{lidar_gs,vggt}, low-cost and efficient scene reconstruction based on sparse point clouds or a limited number of images has become possible.
By further incorporating high-precision ray tracing (RT) simulator, such as Wireless InSite \cite{wireless_insite} and Sionna \cite{Sionna_RT}, into the reconstructed scenes to simulate electromagnetic (EM) wave propagation, one can establish a geometry-centric DT framework.
This natural and practical paradigm offers strong physical interpretability.  
However, its accuracy depends critically on the EM properties assigned to scene objects \cite{DT_emp}. 
Traditional methods typically first classify scene objects and then use empirical lookup tables, such as the ITU recommendations \cite{ITU_R}, to assign them simple and homogeneous material properties as in \cite{itu_rt_boston, itu_rt_wifi}.
Such idealized approximations often overlook the complexity and heterogeneity of real-world materials, and the resulting physical distortion inevitably leads to a sim-to-real gap \cite{DT_bridge}.
Although some studies \cite{calibrate_emp_1,Sionna_Diff_RT,calibrate_emp_2} have attempted to inversely calibrate material properties from channel measurements, this remains a highly nonlinear and complex problem, and precise calibration requires massive and dense measurement data. 
As a result, a fundamental data paradox arises: although DT is intended to alleviate data scarcity, its high-fidelity construction is itself dependent on the large-scale and expensive measurement campaigns.

To bypass the computation and calibration of explicit physics simulators, some studies employ data-driven surrogate models, ranging from kriging interpolation methods to end-to-end generative networks \cite{data_DT_gpr, data_DT_skx, data_DT_diffusion,data_DT_wck, data_DT_bp}. 
These models directly learn the complex mapping from device deployment or system configuration to channel characteristics, and can achieve extremely low inference latency and high fitting accuracy in specific constrained scenarios. 
However, the performance of such data-driven DT-oriented methods depends primarily on the coverage and quality of the training data, rather than on environment-consistent physical representations. 
\cite{data_DT_gpr, data_DT_skx, data_DT_diffusion} train these models using synthetic data generated by RT, thereby avoiding costly real-channel data collection. However, when the EM properties assigned to scene objects are distorted, the trained models inevitably inherit and may even amplify the resulting physical bias. 
\cite{data_DT_wck, data_DT_bp} further introduce a small amount of real measurement data for supervised fine tuning after such pretraining, but they still require a certain amount of data and remain within the paradigm of data fitting. 
Moreover, because the learned mapping is not anchored to explicit propagation physics, these models cannot naturally support new hardware settings, physical-layer designs, or transceiver deployment topologies without retraining. 

The emergence of differentiable scene representation methods in computer vision, such as neural radiance fields (NeRF) \cite{nerf} and 3D Gaussian splatting (3DGS) \cite{3dgs}, has also brought new insights for wireless DT construction.
Methods inspired by NeRF, \cite{nerf2,ris-nerf} adopt neural implicit representations to model continuous radio frequency fields and integrate differentiable propagation and aggregation processes to achieve wireless field rendering based on channel observations.
Similarly, 3DGS-inspired methods \cite{wrf-3dgs,rf-3dgs} use a large number of differentiable Gaussian primitives to represent wireless field distributions, enabling efficient optimization and rendering in specific tasks.
These methods provide flexible scene-level propagation representations and have shown promising performance in various wireless tasks.
However, these methods essentially model the scene-level propagation response holistically and represent the propagation behavior continuously. 
Since the learned representations usually couple the environmental structure, observation coverage, and communication configuration, they often face limited transferability, insufficient supervision, and constrained training stability under inadequate data fitting, especially in sparse measurement conditions.

To overcome the above limitations, this paper proposes a wireless environment digital twin (WEDT) construction paradigm under sparse measurement conditions, aiming to evolve a geometric DT that only describes spatial structure into a wireless environment representation consistent with real propagation observations.
Unlike directly fitting the end-to-end channel mapping or implicitly modeling the entire propagation process, this paper preserves the explicit geometric propagation chains provided by RT and models only the propagation-related EM properties of the environment as a learnable field. 
Through this decomposition, the learning target shifts from the channel response of a specific link to the EM properties of the environment itself.
However,  the limited channel state information (CSI) under sparse measurement conditions is insufficient to directly support reliable calibration of the spatially continuous EM property field. 
To address this issue, this paper further exploits the prior information provided by the geometric DT to construct a Bayesian channel map (BCM) with uncertainty quantification, transforming sparse observations into dense probabilistic supervision and calibrating the EM property field under differentiable channel computation framework.
In this way, the WEDT preserves the physical interpretability of ray tracing while introducing the continuous modeling capability of neural representations and the environmental adaptability of data-driven methods, thereby forming a wireless DT that can represent the real-world environment more effectively.
The main contributions of this paper are summarized as follows:
\begin{itemize}
    \item \textbf{Sparse-measurement WEDT construction:} We propose a WEDT construction paradigm under sparse measurement conditions by formulating it as a reconstructed geometric DT equipped with a calibrated EM property field consistent with real-world propagation. We further characterize an RT-based explicit propagation chain that links geometric propagation paths, the EM property field, and channel observations, thereby enabling channel computation from the calibrated environment representation rather than fitting link-specific channel responses.
    \item \textbf{Geometry-prior Bayesian channel map:} We introduce a geometry-prior BCM to infer dense channel parameters and their uncertainty estimates over the target region. By building a hybrid physical-residual model that combines limited CSI observations with propagation priors extracted from the geometric DT, the BCM provides reliable probabilistic supervision for scene-level EM property field calibration, thereby alleviating the lack of direct supervision under sparse measurements.
    \item \textbf{Uncertainty-aware EM-property field calibration:} We develop an uncertainty-aware calibration method that parameterizes the learnable EM property field as a neural network and embeds it into differentiable RT-based channel computation. The explicit propagation chain provides a differentiable mapping from EM properties to CSI, allowing the network parameters to be optimized under the probabilistic supervision provided by BCM. Meanwhile, the quantified uncertainty adjusts the supervision reliability and ensures stable calibration.
    \item \textbf{Validation of WEDT consistency and utility:} We demonstrate that the calibrated WEDT provides a wireless environment representation that is consistent with real-world propagation. Experimental evaluations of the developed WEDT prototype system in both public and real-world scenes validate this claim, showing that WEDT performs strongly in channel prediction and related tasks. In addition, WEDT shows application potential for wireless tasks such as environment sensing, RIS planning, and AI-oriented data generation.
\end{itemize}

\section{WEDT Formulation}

\subsection{Construction Workflow}

\begin{figure*}[t]
    \centering
    \includegraphics[width=\textwidth]{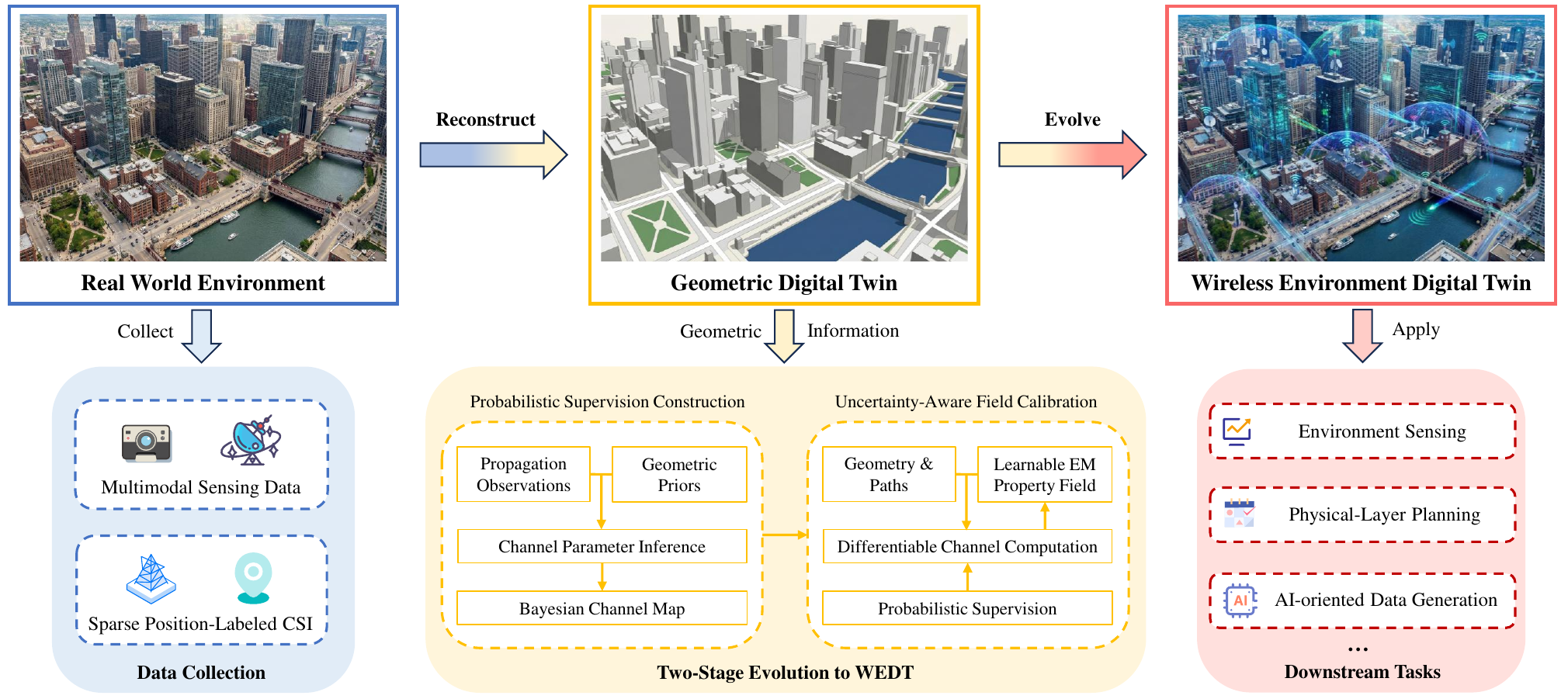}
    \caption{\textbf{Conceptual workflow for constructing a WEDT from a real-world environment.} Multimodal sensing data collected from the real-world environment are used to reconstruct the geometric DT, while sparse position-labeled CSI data provide real propagation observations. The geometric DT is further evolved into the WEDT through a two-stage process. The WEDT has an EM property representation consistent with real-world propagation and can be applied to various downstream tasks.} 
    \label{WEDT_FIG}
    \vspace{-10pt}
\end{figure*}

Fig. \ref{WEDT_FIG} illustrates the proposed conceptual workflow for constructing a WEDT from the real world. 
In this workflow, we first collect two types of data from the scene, where multimodal sensing data such as images and point clouds are used to reconstruct the geometric DT of the environment, and the sparse position-labeled CSI data provide real propagation observations. 
Next, rather than directly fitting the channel responses of the measured links, the workflow aims to utilize
these measurements to calibrate the EM property field attached to the geometric DT.

The main workflow consists of two-stages. 
The first stage converts sparse measurements into dense probabilistic supervision. We combine the propagation prior information extracted from the geometric DT and limited observation samples to infer channel parameters over a wider range of target positions, and quantify the inference uncertainty to evaluate the supervision reliability, yielding the BCM.
The second stage calibrates the EM property field under a differentiable RT-based channel computation framework. For a given transceiver pair, RT on the geometric DT determines the propagation paths and ray-environment interaction points, and the corresponding channel response can then be computed by querying the EM property field at these interaction points. Furthermore, by matching the computed results with the probabilistic supervision provided by the BCM and using uncertainty to adjust the contribution of each supervision sample to the optimization process, we can update and calibrate the learnable field parameters.

Through these two stages, the geometric DT that originally describes only the spatial structure of the scene is equipped with a calibrated EM property field and evolved into a propagation-consistent WEDT. 
As a result, the WEDT can support wireless channel computation that is more consistent with the real environment and can be further applied to downstream tasks such as environmental sensing, physical-layer planning, and AI-oriented data generation.
We define the above WEDT construction workflow as a wireless environment representation and calibration problem under sparse measurements, and formulate it in the next subsection.

\subsection{Problem Formulation}

Building on the above conceptual workflow, we further formulate the WEDT construction problem under sparse measurements.
Given a scene, let $\mathcal{G}$ denote the reconstructed geometric DT of the environment, which contains the scene geometry structures and object surface information used for RT, while the real propagation observations are provided by sparse position-labeled CSI measurement samples, denoted as 
\begin{equation} 
\label{dataset_RE}
{{\mathcal{D}}_{\text{meas}}}=\left\{ \left( {{\mathbf{r}}_{m}},\mathbf{H}_{m}^{\text{meas}} \right) \right\}_{m=1}^{M},
\end{equation}
where ${{\mathbf{r}}_{m}}$ and $\mathbf{H}_{m}^{\text{meas}}$ are the measurement position and CSI of the $m$-th receiver (Rx), respectively. Under the sparse acquisition setting considered in this paper, the observation perspective is limited by a fixed transmitter (Tx), while the size $M$ of Rx samples with accurate location labels is limited and their spatial distribution is uneven.
Therefore, directly using ${{\mathcal{D}}_{\text{meas}}}$ to calibrate the continuous and complex EM property distribution in the environment is an under-supervised and ill-posed problem.

Based on the reconstructed geometric DT $\mathcal{G}$, the WEDT further needs to learn a representation related to real wireless propagation, which is a spatially continuous effective EM property field defined on the scene surfaces. For any location $\mathbf{q}$ on the scene surface, this field is expressed as 
\begin{equation}
{{\mathcal{M}}_{\Theta }}( \mathbf{q} )=\{{{\epsilon }_{r}}( \mathbf{q} ),\sigma ( \mathbf{q} ),S( \mathbf{q} ),{{K}_{\chi }}( \mathbf{q} )\},
\end{equation}
where ${{\epsilon}_{r}}$, $\sigma$, $S$, and ${{K}_{\chi}}$ denote the relative permittivity, conductivity, scattering coefficient, and cross-polarization discrimination factor, respectively. $\Theta$ is the set of trainable parameters of the EM property field.
Unlike traditional EM property assignment methods based on object categories and empirical lookup tables, ${{\mathcal{M}}_{\Theta }}( \mathbf{q} )$ is learnable and spatially continuous, enabling it to adapt to different scenes and characterize microscopic variations in EM properties.
In the implementation, ${{\mathcal{M}}_{\Theta}}$ is represented by a parameterized neural network, as detailed in Section III-B.

To alleviate the insufficient supervision caused by sparse measurements, we introduce an intermediate probabilistic supervision set called the BCM, expressed as 
\begin{equation}
\label{BCM_definition}
{{\mathcal{D}}_{\text{B}}}=\{(\mathbf{r}_{m}^{*},{{\bm{\hat{\alpha }}}_{m}},{{\bm{\kappa }}_{m}})\}_{m=1}^{{{M}^{*}}},\quad {{M}^{*}}\gg M, 
\end{equation}
where $\mathbf{r}_{m}^{*}$ denotes the target Rx location, ${{\bm{\hat{\alpha}}}_{m}}$ represents the channel parameters inferred from sparse measurements and geometric priors, ${{\bm{\kappa}}_{m}}$ is the corresponding uncertainty estimate, and ${{M}^{*}}$ denotes the size of the probabilistic supervision set.
The specific construction method will be introduced in Section III-A. 
It should be emphasized that ${{\mathcal{D}}_{\text{B}}}$ is not the final environment representation of the WEDT, but an intermediate probabilistic supervision with uncertainty characterization for calibrating ${{\mathcal{M}}_{\Theta}}$. 
The final WEDT representation remains the geometric DT together with the calibrated EM property field.

Therefore, the WEDT construction under sparse measurements can be formulated as
\begin{equation}
\label{WEDT_formulation}
{{\Theta }^{\star }}=\argmin_{\Theta } \,{{\mathcal{L}}_{\text{u}}}\left( \mathcal{C}(\mathcal{G},{{\mathcal{M}}_{\Theta }}),{{\mathcal{D}}_{\text{B}}} \right), 
\end{equation}
where $\mathcal{C}(\mathcal{G},{{\mathcal{M}}_{\Theta }})$ denotes the RT-based channel computation process jointly determined by the geometric DT and the EM property field, including CSI computation and the extraction of channel parameters for supervision.
${{\mathcal{L}}_{\text{u}}}$ denotes the uncertainty-aware calibration loss that accounts for the reliability of probabilistic supervision.
This formalization indicates that the core of WEDT construction is not to directly fit the link-level channel mapping, but to calibrate the scene-level EM property representation within a physical channel computation framework.
The above optimization relies on a key fact that once the geometric propagation paths are determined, the computed channel response is a differentiable function of the EM property field. The next subsection explains this dependency.

\subsection{EM-Property-Field-Dependent Channel Computation}

The WEDT formulation in (\ref{WEDT_formulation}) requires a channel computation model that maps the EM property field attached to the geometric DT to CSI, and this subsection clarifies this forward dependence. 
In general, once the geometric DT and transceiver positions are determined, RT can identify the propagation paths and their geometric parameters. 
The remaining learnable scene-dependent component is the EM property fields queried at the ray-environment interaction points, which affect the final CSI through field interactions along each path.

Given an OFDM transceiver with carrier frequency $f_c$, operating bandwidth $W$, $N$ subcarriers, and $L$ paths, the CSI of the $n$-th subcarrier can be expressed as 
\begin{equation} 
H[ n ]=\sum\nolimits_{i=1}^{L}{{{a}_{i}}{{\text{e}}^{-\text{j}2\pi {{f}_{n}}{{\tau }_{i}}}}},\label{CSI_n}
\end{equation}
where $a_i$ and $\tau_i$ denote the complex channel coefficient and the delay of the $i$-th path, respectively. $f_n = f_c + n\Delta f$ denotes the frequency of the $n$-th subcarrier,and $\Delta f = W/N$ is the subcarrier spacing.

Based on RT, we can identify the propagation paths between the transmitter and receiver. 
The path delay ${\tau}_{i}$ can be directly calculated from the total path propagation distance $d_{i}^{\text{total}}$ and the speed of light $c$ as $\tau_i = d_{i}^{\text{total}}/c$, while the complex channel coefficient $a_i$ captures antenna characteristics and propagation effects and is modeled as \cite{RT_channel_model}
\begin{equation}
{{a}_{i}}=\mathbf{c}_{R}^{H}\left( \phi _{i}^{\text{Rx}},\theta _{i}^{\text{Rx}} \right){{\mathbf{A}}_{i}}{{\mathbf{c}}_{T}}\left( \phi _{i}^{\text{Tx}},\theta _{i}^{\text{Tx}} \right),\label{a}
\end{equation}
where $(\phi_{i}^{\text{Tx}}, \theta_{i}^{\text{Tx}})$ and $(\phi_{i}^{\text{Rx}}, \theta_{i}^{\text{Rx}})$ denote the departure and arrival angles of the $i$-th path respectively. 
$\mathbf{c}_{T}(\cdot) \in \mathbb{C}^{2\times 1}$ and $\mathbf{c}_{R}(\cdot) \in \mathbb{C}^{2\times 1}$ represent the polarized field response vectors of the Tx and Rx antenna. 
The propagation matrix $\mathbf{A}_i \in \mathbb{C}^{2\times 2}$ characterizes the cumulative effect of all physical interactions experienced by the $i$-th path in the environment.

\begin{figure}[!t]
  \centering
  \includegraphics[width=\linewidth]{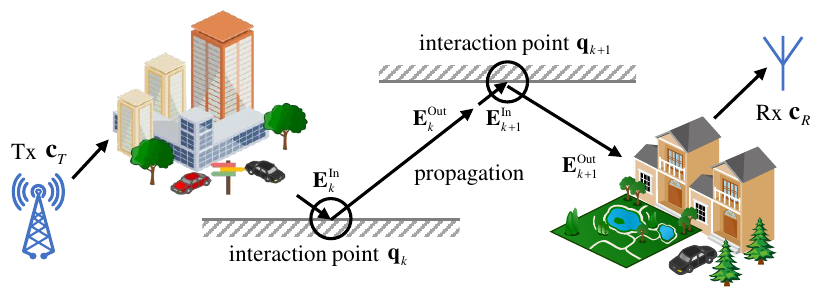}
  \caption{\textbf{Path propagation and interaction}. The wireless signal is radiated by the Tx, undergoes $K$ interactions in the environment, and is finally captured by the Rx.}
  \label{Path}
  \vspace{-10pt}
\end{figure}

As shown in Fig. \ref{Path}, we consider a propagation path that undergoes $K$ interactions in the environment, and track the propagation of its polarized electric field along this path.
Specifically, the Tx first radiates the initial electric field into the environment. 
When the path propagates to the $k$-th interaction point ${\mathbf{q}}_{k}$, the corresponding incident electric field $\mathbf{E}_{k}^{\text{In}}$ is obtained through coordinate transformation, which then interacts with the environment to form the outgoing electric field $\mathbf{E}_{k}^{\text{Out}}$ directed towards the next interaction point ${{\mathbf{q}}_{k+1}}$. 
After undergoing all interactions, the Rx captures the final incident electric field of this path.

For the antenna polarized field response vectors, we adopt a dual-polarized antenna model and define it on the orthogonal horizontal and vertical polarization bases
\begin{equation}
\label{antenna}
\mathbf{c}( \phi ,\theta  )={{10}^{G( \phi ,\theta  )/20}}{{[ \cos ( \zeta  ),\sin ( \zeta  ) ]}^{T}},
\end{equation}
where $\zeta$ is the polarization slant angle. $G(\phi, \theta)$ denotes the antenna radiation pattern in dB, which can be further modeled according to the 3GPP specification as \cite{3gpp38901}
\begin{equation}
\label{antenna_gain}
G( \phi ,\theta )=G_{\max }^{{}}-\min \left\{ 12\left( {{{\bar{\phi }}}^{2}}+{{{\bar{\theta }}}^{2}} \right),\mathcal{A}_{\max }^{{}} \right\},
\end{equation}
where $G_{\max}$ denotes the maximum directional gain, $\mathcal{A}_{\max}$ represents the maximum front-to-back attenuation, $\bar{\phi} = \phi/\phi_{3\text{dB}}$ and $\bar{\theta} = (\theta - \pi/2)/\theta_{3\text{dB}}$ are the antenna's normalized azimuth and elevation angles defined by the corresponding half-power beamwidths $\phi_{3\text{dB}}$ and $\theta_{3\text{dB}}$.

The above propagation process of the electric field along each segment can be characterized by the field interaction matrix $\mathbf{F}$ and the basis transformation matrix $\mathbf{T}$, and after accumulating the interactions of all segments along the path, the path propagation matrix $\mathbf{A}$ can be derived as \cite{Sionna_Diff_RT}
\begin{equation}
\label{F&T}
\mathbf{E}_{k}^{\text{Out}}={{\mathbf{F}}_{k}}\mathbf{E}_{k}^{\text{In}},\text{ }\mathbf{E}_{k+1}^{\text{In}}={{\mathbf{T}}_{k+1}}\mathbf{E}_{k}^{\text{Out}}, \dots.,
\end{equation}
\begin{equation}
\label{A}
\mathbf{A}=\frac{c}{4\pi {f}_{c}}\prod\nolimits_{k=0}^{K}{{{\mathbf{T}}_{k+1}}\mathbf{F}_{k}}.
\end{equation}
The $\mathbf{T}_{k+1} \in \mathbb{C}^{2 \times 2}$ ensures the correct mapping of the electric field vector between local coordinate systems.
While the ${{\mathbf{F}}_{k}}\in{{\mathbb{C}}^{2\times 2}}$ models the specific interaction of the electric field at ${{\mathbf{q}}_{k}}$ and its subsequent propagation attenuation.
When $k=0$, $\mathbf{F}_{0}$ describes the free-space propagation of the electric field from the antenna to the first interaction point $\mathbf{q}_{1}$. 
For $1\le k\le K$, ${{\mathbf{F}}_{k}}$ depends on the EM property field ${{\mathcal{M}}_{\Theta }}({{\mathbf{q}}_{k}})$ at the interaction point, while the interaction type and local geometric quantities are determined by the RT on the geometric DT, which can be written as 
\begin{equation}
{{\mathbf{F}}_{k}}=\mathcal{I}_{k}^{\mathcal{G}}\left( {{\mathcal{M}}_{\Theta }}({{\mathbf{q}}_{k}}) \right),
\end{equation}
where $\mathcal{I}_{k}^{\mathcal{G}}(\cdot )$ is a path-dependent interaction operator determined by the geometric DT $\mathcal{G}$ and the $k$-th ray interaction state, covering propagation mechanisms such as coherent specular reflection, diffraction, and incoherent diffuse scattering, with detailed formulations available in \cite{spreading_factor, ITU-R-P526-15, scattering, Sionna_report}. 

Therefore, for the reconstructed geometric DT and a given transceiver pair, RT first determines the paths, interaction points, and interaction types. 
Conditioned on these geometric quantities, the EM property fields affect the path propagation matrix through the field interaction matrices at each interaction point, thereby determining the complex channel coefficients of each path and the resulting CSI. This gives the explicit dependency chain as 
\begin{equation}
\mathcal{M}_{\Theta}(\mathbf{q}_k) \to \mathbf{F}_k \to \mathbf{A}_i \to a_i \to H[n] \to \mathbf{H}.
\end{equation}
This dependency chain establishes the connection between the EM property fields and the CSI.
Once the geometric paths are determined, representing ${{\mathcal{M}}_{\Theta}}({{\mathbf{q}}_{k}})$ as a differentiable parameterized function makes the RT-based channel computation differentiable with respect to $\Theta$. Therefore, we can further utilize CSI-related supervision to optimize the EM property fields through the explicit propagation process.
We will detail the specific implementation of this calibration in Section III.

\section{Two-Stage WEDT Calibration Under Sparse Measurements}

We formulated WEDT construction as the calibration of the scene-level EM property field attached to the geometric DT in Section II. 
To make this calibration feasible under sparse CSI measurements, this section proposes a two-stage implementation method. 
We first construct dense probabilistic supervision by inferring stable channel parameters and their uncertainty in the target region based on geometric priors. 
Then, we embed the learnable EM property field into differentiable channel computation based on RT and calibrate it using uncertainty-aware supervision. 
As a result, sparse measurements are used to calibrate the EM property field through an intermediate probabilistic supervision layer, allowing the geometric DT to evolve toward a propagation-consistent WEDT.

\subsection{Geometry-Prior Probabilistic Supervision Construction} 
Before calibrating the EM property fields, the limited measured propagation observations need to be converted into supervisory signals for stable optimization, and this subsection constructs a BCM based on the geometric DT prior to achieve this goal. 
This BCM can provide inferred channel parameters and their uncertainty estimates within the target area, thereby forming a probabilistic supervision bridge between sparse position-labeled CSI measurements and the subsequent calibration of the EM property fields.

Directly using the measured dataset ${{\mathcal{D}}_{\text{meas}}}$ to calibrate the EM property fields is difficult. 
On the one hand, the sparsity of ${{\mathcal{D}}_{\text{meas}}}$ prevents it from providing sufficient supervision. 
On the other hand, although the raw CSI contains complete channel information, its high dimensionality and extreme sensitivity to wavelength-scale displacement can cause severe data fluctuations.
Therefore, instead of directly interpolating the raw CSI, we extract more stable and physically meaningful channel parameters from it for inference within the target area. 
The following construction process applies to a general channel parameter $\alpha$.
In the implementation, we select the logarithmic channel gain $p$ and the root-mean-square (RMS) delay spread $\tau$, meaning $\alpha \in \{p,\tau \}$, as they respectively characterize the key physical properties of energy attenuation and time dispersion in wireless channels.

Although channel parameters are more stable than the raw CSI, inference based only on position information alone is insufficient. 
In complex real-world environments, two spatially adjacent Rx may have substantially different channel responses due to blockage, reflection, and scattering. 
Conversely, two distant Rx may exhibit similar channel responses if they have similar propagation path structures. 
This indicates that the correlation between channel parameters is not determined only by position, but also depends on the propagation conditions induced by the environmental geometry.

To characterize such propagation differences, we exploit the prior information contained in the reconstructed geometric DT. 
Although the geometric DT does not contain electromagnetically consistent propagation properties, it provides reliable spatial structures and geometric propagation information. 
Therefore, we utilize RT in the geometric DT to extract the geometric propagation features for each Rx position and fuse them with the position features to obtain the fused feature vector as 
\begin{equation}
\bm{\omega }(\mathbf{r})=[\mathbf{r},{{\bm{\varphi }}_{\text{path}}}(\mathbf{r}),{{\bm{\varphi }}_{\text{fade}}}(\mathbf{r})]
\end{equation}
where ${{\bm{\varphi}}_{\text{path}}}$ denotes the ray path structure feature, which characterizes the richness of propagation paths. It reflects the scatterer distribution around the Rx and the environmental structural complexity by analyzing the existence of the LOS path, the number of traced paths, and the interaction orders of multipath components.
${{\bm{\varphi}}_{\text{fade}}}$ represents the spatial fading statistical feature, which provides a statistical description of local channel variations. Based on the frequency-space duality principle \cite{frequency_spacing}, it uses RT to simulate the broadband frequency-domain samples of the Rx to approximate local spatial perturbations, thereby characterizing the relative severity of small-scale fading in the current region.

Based on the fused features enhanced by the above geometric prior information, we construct a hybrid physical-residual model for each channel parameter $\alpha$. 
Its core idea is to decompose the channel parameter into a macroscopic physical trend ${{g}_{\alpha}}$ and an environment-dependent residual ${{\xi }_{\alpha}}$.

The macroscopic physical model ${{g}_{\alpha}}$ aims to capture the baseline distance-dependent spatial trend of the channel parameters and provides a physically interpretable anchor for the model. 
Based on classic propagation models, the logarithmic channel gain model ${{g}_{p}}$ \cite{Path_loss} and the power-law RMS delay spread model ${{g}_{\tau}}$ \cite{Delay_spread} can be modeled as
\begin{equation}
{{g}_{p}}(\mathbf{r})={{G}_{0}}-10{{n}_{0}}{{\log }_{10}}d(\mathbf{r}),\text{  }{{g}_{\tau }}(\mathbf{r})={{A}_{0}}d{{(\mathbf{r})}^{{{b}_{0}}}},
\end{equation}
where $d(\mathbf{r})$ is the distance between the Tx and Rx, and $\left\{ {{G}_{0}},{{n}_{0}},{{A}_{0}},{{b}_{0}} \right\}$ are the model parameters that can be learned from the measurement samples.

However, directly using all measurement samples to train the above propagation models is unreliable. 
This is because these propagation models are essentially statistical models, while the measured samples are spatially sparse and thus difficult to fully reflect the global statistical distribution of the channel. 
A few severely blocked samples may also distort the distance-dependent trend, leading to biased model parameters.
To address this, we propose a geometry-constrained model training strategy. 
Specifically, we utilize the ray path structure feature ${{\bm{\varphi}}_{\text{path}}}$ extracted from the geometric DT to select samples with LOS propagation characteristics, and train the model parameters of ${{g}_{\alpha }}$ on this subset using the least squares. 
This strategy ensures that ${{g}_{\alpha }}$ can provide a relatively stable propagation baseline.

Given that the residual component of the channel parameters is highly dependent on complex environmental structures and usually lacks an explicit analytical form, we use Gaussian process regression (GPR) to model the residual ${{\xi}_{\alpha }}$. 
GPR is a nonparametric Bayesian method that is not only suitable for nonlinear function inference with limited samples, but also provides uncertainty estimates for the inferred results \cite{GPR}.
Since ${{g}_{\alpha}}$ has captured the baseline trend of the channel parameters, we adopt a zero-mean Gaussian process prior. 
Compared with traditional kernel functions that rely only on positional features \cite{GPR_pos}, we further define the kernel function in the fused feature space enhanced by geometric priors, so that the residual correlation depends on both spatial proximity and propagation structure similarity. 
For any two fused features, the kernel function of ${{\xi}_{\alpha}}$ is defined as 
\begin{equation}
{{k}_{\alpha }}(\bm{\tilde{\omega }},\bm{\tilde{{\omega }}}')={{k}_{f}}(\bm{\tilde{\omega }},\bm{\tilde{{\omega }}}')+{{k}_{n}}(\bm{\tilde{\omega }},\bm{\tilde{{\omega }}}'),
\end{equation}
where $\bm{\tilde{\omega}}$ is the normalized fused feature after eliminating internal dimensional differences. 
${{k}_{f}}$ is the fused feature kernel. In this paper, we adopt the Matérn kernel with automatic relevance determination to balance the smoothness and local irregularity of the environmental structure, and to adaptively determine the importance of different feature dimensions. 
${{k}_{n}}$ is the white noise kernel, which characterizes measurement noise and unexplained perturbations.

During training, we first remove the macroscopic physical trend from the measured channel parameters to obtain the residual training samples. 
Then, the hyperparameters of the kernel function are trained by maximizing the log marginal likelihood. 
For ${{\mathbf{\tilde{\omega}}}^{*}}$ corresponding to the position to be inferred, the GPR posterior can provide the mean and variance of the channel parameter residual ${{\xi}_{\alpha}}$ as 
\begin{align}
\mathbb{E}[{{\xi }_{\alpha }}({{\bm{\tilde{\omega }}}^{*}})]&=\mathbf{k}_{f^{*}}^{T}{{({{\mathbf{K}}_{f}}+\sigma _{n}^{2}\mathbf{I})}^{-1}}\mathbf{y}_{\alpha }^{res}, 
 \label{xi_pred}\\
\text{Var}[{{\xi }_{\alpha }}({{\bm{\tilde{\omega }}}^{*}})]&={{k}_{\alpha^{**}}}-\mathbf{k}_{f^{*}}^{T}{{({{\mathbf{K}}_{f}}+\sigma _{n}^{2}\mathbf{I})}^{-1}}\mathbf{k}_{f^{*}}, \label{xi_confidence}
\end{align}
where $\mathbf{y}_{\alpha}^{res}$ is the residual vector of the measured channel parameter $\alpha$. ${{\mathbf{k}}_{{{f}^{*}}}}$ and ${{\mathbf{K}}_{f}}$ respectively represent the covariance vector between the inference point and the training set as well as the auto-covariance matrix of the training set. ${{k}_{\alpha^{**}}}$ is the variance of the inference point itself.

Thus, once the hybrid physical-residual model training is completed, we can query the model at a target position ${{\mathbf{r}}^{*}}$ to obtain the inferred value and uncertainty of the channel parameter $\alpha$ as 
\begin{align}
\hat{\alpha }({{\mathbf{r}}^{*}})&={{g}_{\alpha }}({{\mathbf{r}}^{*}})+\mathbb{E}[{{\xi }_{\alpha }}(\bm{\tilde{\omega }}({{\mathbf{r}}^{*}}))],\\
{{\kappa }_{\alpha }}({{\mathbf{r}}^{*}})&=\text{Var}[{{\xi }_{\alpha }}(\bm{\tilde{\omega }}({{\mathbf{r}}^{*}}))],
\end{align}
where $\hat{\alpha}$ is the sum of the macroscopic physical model and the posterior mean of the residual model, and ${{\kappa}_{\alpha}}$ is the posterior variance of the residual model. 
Thus, performing the above inference for all target positions $\{\mathbf{r}_{m}^{*}\}_{m=1}^{{{M}^{*}}}$ and all selected channel parameters $\alpha \in \{p,\tau \}$, we can obtain the BCM defined in (\ref{BCM_definition}).
This BCM unifies the geometric propagation priors, sparse real measurements, and uncertainty estimates into dense probabilistic supervision. 
The next subsection further embeds this supervision into the differentiable RT-based channel computation framework to achieve closed-loop calibration of the EM property fields.

\subsection{Uncertainty-Aware EM Property Field Calibration} 

The BCM constructed in the previous subsection provides dense inferred channel parameters and their uncertainty estimates for the target region in the environment. 
However, the BCM remains a location-level channel parameter inference model and is not the final representation of the proposed WEDT. 
This subsection aims to convert the probabilistic supervision provided by the BCM into the calibration of the EM property field through an explicit differentiable propagation chain. 
The calibrated EM property field and the geometric DT jointly form the core representation of the WEDT, enabling it to support wireless channel computation consistent with the real environment and facilitate subsequent wireless tasks.

\begin{figure}[!t]
  \centering
  \includegraphics[width=\linewidth]{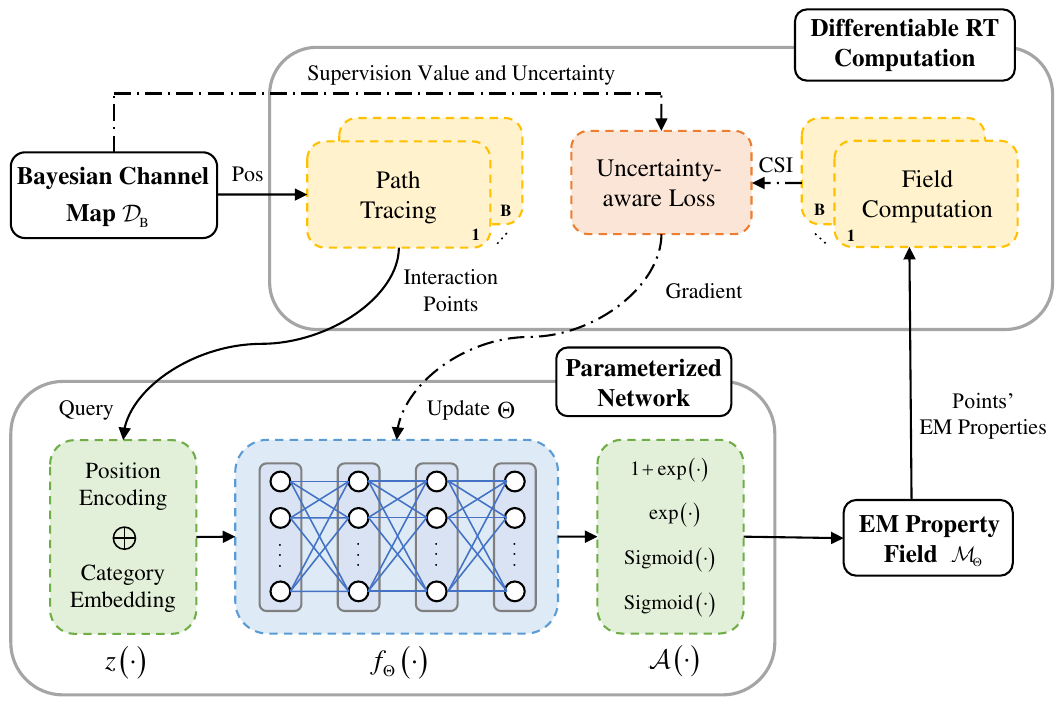}
  \caption{\textbf{ Uncertainty-aware EM property field calibration framework based on differentiable RT.} The BCM provides Rx samples, inferred channel parameters, and uncertainty estimates for training. After calibration, the probabilistic supervision and gradient feedback represented by the dashed branch are removed, and the remaining forward pipeline is used for CSI computation in the WEDT.}
  \label{calibration}
  \vspace{-10pt}
\end{figure}

As shown in Fig. \ref{calibration}, we establish an uncertainty-aware EM property field calibration framework based on differentiable RT. 
Specifically, for each target Rx sample position in the BCM, the RT first determines the propagation paths in the geometric DT and identifies the ray-environment interaction points. 
Then, a parameterized neural network takes the spatial positions of these interaction points as input to query the corresponding EM property field.
Given the traced propagation paths and the EM property field, differentiable channel computation performs field calculation to obtain the corresponding CSI and extracts the channel parameters for supervision. 
The computed results are then matched with the reference inferred values provided by the BCM, while the uncertainty associated with the BCM is used to adjust the reliability of each supervision sample. 
According to the explicit dependency chain established in Section II-C, this process is differentiable with respect to the parameters of the EM property field, so the loss gradients can be backpropagated to the parameterized network, thereby calibrating the EM property field.

Thus, the sparse measured data are not directly used to fit link-level channel responses, but instead calibrate the EM properties at the environment level through the constructed BCM and the explicit propagation chain. 
In addition, after the EM property field is calibrated, the dashed branch in Fig. 3 that represents supervision and gradient feedback can be removed, and the BCM can be replaced by any new transceiver pair, with the remaining forward chain serving as the procedure for computing its corresponding channel in the WEDT.

For each ray-environment interaction point $\mathbf{q}$ obtained via path tracing, the corresponding EM property field is parameterized as
\begin{equation} 
\label{F_EMPNN}
{{\mathcal{M}}_{\Theta }}( \mathbf{q} )=\mathcal{A}( {{f}_{\Theta }}( z( \mathbf{q} ) ) ).
\end{equation}
Here, $z(\cdot)$ denotes the network input encoding. 
Since directly using low-dimensional position coordinates as input makes it difficult for the network to capture the high-frequency variations and texture details of the environmental EM property distribution, we normalize the coordinates and map them into a Fourier feature encoding vector \cite{nerf}. 
By default, $z(\cdot)$ is this coordinate encoding. 
For scenarios with object category priors, a learnable category embedding can be further introduced and concatenated with the coordinate encoding as the network input.
$f_{\Theta}(\cdot)$ is a backbone neural network with trainable parameters $\Theta$, designed with hidden fully connected layers followed by ReLU activation functions and ending with a four-dimensional linear output layer. 
$\mathcal{A}(\cdot)$ denotes a specific activation function mapping the backbone network output into a valid range, thereby strictly satisfying the physical constraints of the EM properties. For the distinct value ranges of $\epsilon_{r} \ge 1$, $\sigma \ge 0$, $S \in [0,1]$, and $K_{\chi} \in [0,1]$, the respective formulations are given by
\begin{align}
{{\epsilon }_{r}}&=1+\exp ( \cdot ),
~~
\sigma =\exp ( \cdot ), \\
S &=\text{sigmoid}( \cdot ),
{{K}_{\chi }}=\text{sigmoid}( \cdot ),
\end{align}
where $\text{sigmoid}( x)=1/\left( 1+\exp ( -x ) \right)$.

The channel parameters $p$ and $\tau$ extracted from the CSI for supervision are expressed as
\begin{align}
p&=10{{\log }_{10}}{{p}_{lin}},\text{ }{{p}_{lin}}=\sum\nolimits_{\ell }{{{| h[ \ell ] |}^{2}}}, \\
\tau &=\sqrt{{{\sum\nolimits_{\ell }{\left( \frac{\ell -\tilde{\tau }}{W} \right)}}^{2}}\frac{{{| h[ \ell  ] |}^{2}}}{{{p}_{lin}}}},
\end{align}
where $h[\ell]=\frac{1}{\sqrt{N}}\sum\nolimits_{n=-N/2}^{N/2-1}{H[n]\exp \left( \text{j}\frac{2\pi }{N}n\ell \right)}$ with $\ell =-\frac{N}{2},\ldots ,\frac{N}{2}-1$ denotes the discrete time domain Channel Impulse Response (CIR) obtained via the IDFT of CSI. $\tilde{\tau}=\sum\nolimits_{\ell }{\ell {{| h[ \ell ] |}^{2}}}/ {{p}_\text{lin}}$ represents the average delay. $p_\text{lin}$ is the linear channel gain.

Since the BCM provides probabilistic inference results with uncertainty for each supervised quantity rather than deterministic labels, we optimize the network parameters by minimizing an uncertainty-aware negative log-likelihood loss. 
This encourages the channel parameters extracted from CSI computation under the corresponding EM property field to have higher likelihood under the target distributions provided by the BCM. 
The uncertainty-aware loss function for calibrating the parameters of the EM property field is defined as 
\begin{align}
  & {{\mathcal{L}}_{u}}=-\log \left( \Pr \left( \left\{ \mathbf{\alpha }_{i}^{\text{RT}}(\Theta ) \right\}_{i=1}^{B}|\left\{ {{{\mathbf{\hat{\alpha }}}}_{i}},{{\mathbf{\kappa }}_{i}} \right\}_{i=1}^{B} \right) \right) \notag \\ 
 & =\frac{1}{B}\sum\limits_{i=1}^{B}{\sum\limits_{\alpha \in \{p,\tau \}}{{{\eta }_{\alpha }}}}\left( \frac{{{(\alpha _{i}^{\text{RT}}(\Theta )-{{{\hat{\alpha }}}_{i}})}^{2}}}{2{{\kappa }_{\alpha ,i}}}+\frac{1}{2}\log \left( 2\pi {{\kappa }_{\alpha ,i}} \right) \right),  
\end{align}
where $B$ is the batch size, and $\eta_{\alpha}$ is an adjustment parameter that balances the dimensional differences and gradient contributions of different channel parameters. 
$\alpha^{\text{RT}}(\Theta)$ denotes the channel parameter extracted by differentiable RT under the current EM property field parameter $\Theta$. 
This loss function assigns larger weights to supervision samples with lower uncertainty and reduces the impact of highly uncertain supervision samples on parameter updates. 
Therefore, the calibration process can adaptively adjust the supervision strength according to the regional reliability provided by the BCM, thereby improving the stability of EM property field parameter learning under sparse measurements.

In practice, due to system-level factors such as hardware calibration, antenna normalization and link budget mismatch, the channel gain computed by RT may exhibit an almost constant power offset relative to the actual measured value, which should not be learned by the EM property field since it is not caused by scenario-dependent EM properties. 
To distinguish this global mismatch from environment-related calibration, we introduce a scalar bias term $\beta \in \mathbb{R}$ for the logarithmic channel gain. 
At training iteration $t$, its instantaneous estimate within the current batch is calculated as
\begin{equation}
\label{instantaneous_beta}
{{\hat{\beta }}_{t}}=\argmin_{\beta} \,\frac{1}{B}\sum\limits_{i=1}^{B}{{{\left( \left( p_{i}^{\text{RT}}( \Theta )+\beta  \right)-{{{\hat{p}}}_{i}} \right)}^{2}}}.
\end{equation}
Then, we adopt an exponential moving average strategy to update the bias parameter as
\begin{equation}
\label{beta_update}
{{\beta }_{t}}=\lambda {{\beta }_{t-1}}+(1-\lambda ){{\hat{\beta }}_{t}},
\end{equation}
where $\lambda$ is the tuning factor. 
The updated $\beta$ decouples the fixed power offset from the calibration of the EM property field, preventing the parameterized network from being overoptimized to compensate for system-level mismatch.

In summary, this subsection converts the probabilistic supervision with uncertainty provided by the BCM into the calibration of the environmental EM property field through an explicit differentiable propagation chain. 
In this way, the WEDT no longer relies on link-level channel fitting, but instead performs channel computation based on the geometric DT and the calibrated EM property field, thereby forming a wireless environment representation with stronger physical consistency.

\section{Experimental Setup}

Having presented the WEDT formulation and its two-stage calibration method, we now describe the experimental setup used to assess both the propagation consistency and the practical utility of the proposed framework. 
This section introduces the considered scenes, the data acquisition process, the training configuration, and the baselines and evaluation metrics used in the subsequent experiments.

\begin{figure}[!t]
  \centering
  \includegraphics[width=\linewidth]{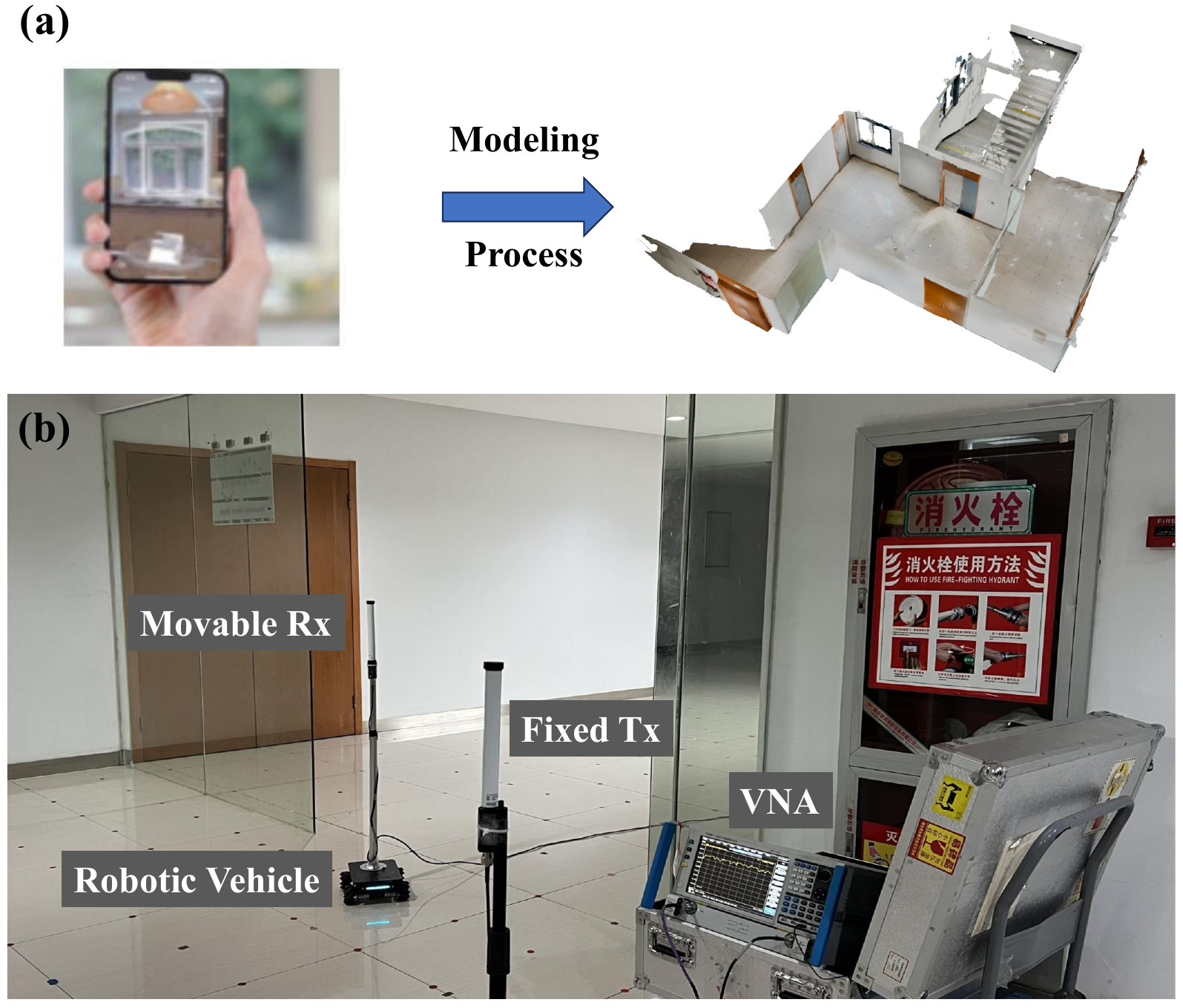}
  \caption{\textbf{Real-world scene and measurement.} (a) Geometric DT reconstruction via smartphone. (b) Position-labeled CSI measurement platform with a fixed Tx, a mobile Rx on a robotic vehicle, and a VNA.}
  \label{Real_Scenario}
  \vspace{-10pt}
\end{figure}

\subsection{Scenarios and Data Acquisition} 
We developed the WEDT prototype system and evaluated its performance in two public scenes \cite{scene1,scene2} and one real-world scene. 
These scenes cover both indoor and outdoor environments, and operate in the WiFi 2.6 GHz, fifth-generation commercial network 3.5 GHz, and U6G 6.4 GHz bands, respectively. 
For the two public scenes, we directly used the provided 3D scene models, while the real-world scene is a corridor environment located in Wireless Valley, Nanjing, China. 
As shown in \ref{Real_Scenario}(a), using a smartphone equipped with a camera and LiDAR, we completed the geometric DT reconstruction of the real-world scene within a few minutes through the 3D Scanner App \cite{3dscannerapp}.

For each scene, we then constructed a position-labeled CSI dataset covering the entire scene plane under a fixed Tx setting. 
All scenes adopt the OFDM signal configuration with $W=200\text{MHz}$ and $N=256$.
In the public scenes, the complete material annotation information is not fully available. Therefore, we first classify the objects in the scenes, then assign reference EM properties to them according to the ITU recommendations \cite{ITU_R}, and use Sionna RT to generate the corresponding reference CSI dataset.
In the real-world scene, we built a measurement platform as shown in Fig. 4(b), which consists of a fixed Tx, a mobile Rx mounted on a robotic vehicle, and a vector network analyzer (VNA). The VNA measures and records the CSI in real time as the Rx moves to different positions.
The Tx and Rx used in measurements employ the same omnidirectional antenna with parameters ${{\phi}_{3\text{dB}}}={{360}^{\circ }}$, ${{\theta}_{3\text{dB}}}={{17}^{\circ }}$, $G_{\max}^{{}}=8\text{dBi}$, and $\mathcal{A}_{\max}^{{}}=30\text{dBi}$. For consistency, the same antenna configuration is also adopted in the experiments on the public scenes.

\subsection{Implementation and Training Configuration}
In each scene, the WEDT is trained on a sparse subset randomly sampled from the full dataset. 
Unless otherwise specified, the number of training samples is set to $M=30$, which accounts for less than 3\% of the full dataset. 
This sparse subset is first used to train the corresponding hybrid physical-residual channel parameter inference model, and the BCM is constructed in the same spatial region as the full dataset. 
The obtained BCM provides dense probabilistic supervision for the subsequent calibration of the EM property field. 
The calibration of the EM property field is completed by embedding it into differentiable channel computation implemented by Sionna RT. 
By default, the parameterized network only takes the Fourier feature encoding of the interaction point positions as input and does not use object class priors.

We train the model on an NVIDIA V100 GPU with 32 GB memory. During training, the Adam optimizer is used with a batch size of 32 and a learning rate of 1e-3, and each scene is trained for 10k iterations.
To accelerate training, all ray paths are precomputed and cached before training, so path tracing is not required in each iteration.

\subsection{Baselines and Metrics}
We compare WEDT with three baselines: 
\begin{itemize}
    \item \textbf{Differentiable RayTracing (Diff-RT) \cite{Sionna_Diff_RT}:} Diff-RT directly optimizes the material parameters within the differentiable ray tracing framework.
    \item \textbf{Neural Radio-Frequency Radiance Fields (NeRF$^{\bm{2}}$) \cite{nerf2}:} NeRF$^2$ learns the implicit propagation field from channel observation data.
    \item \textbf{Gaussian Process With Positional Uncertainty (GP-PU) \cite{GPR_pos}:} GP-PU performs interpolation based on Gaussian processes and measurement position uncertainty.
\end{itemize}

We adopt three evaluation metrics:
(1) Mean absolute logarithmic error (MALE), which quantifies the absolute numerical error in the predicted logarithmic channel gain; 
(2) Structural similarity index measure (SSIM), which evaluates the similarity of the predicted channel gain maps in terms of brightness, contrast, and structure; 
(3) Mean cosine similarity (MCS), which measures the directional consistency of the predicted CSI vectors.

\section{Results and Application Studies}

Based on the above experimental setup, we next systematically evaluate the overall performance of the proposed WEDT under sparse measurements and further investigate its application in downstream wireless tasks.

\subsection{Performance Evaluation}

\begin{figure*}[t] 
    \centering
    \includegraphics[width=\textwidth]{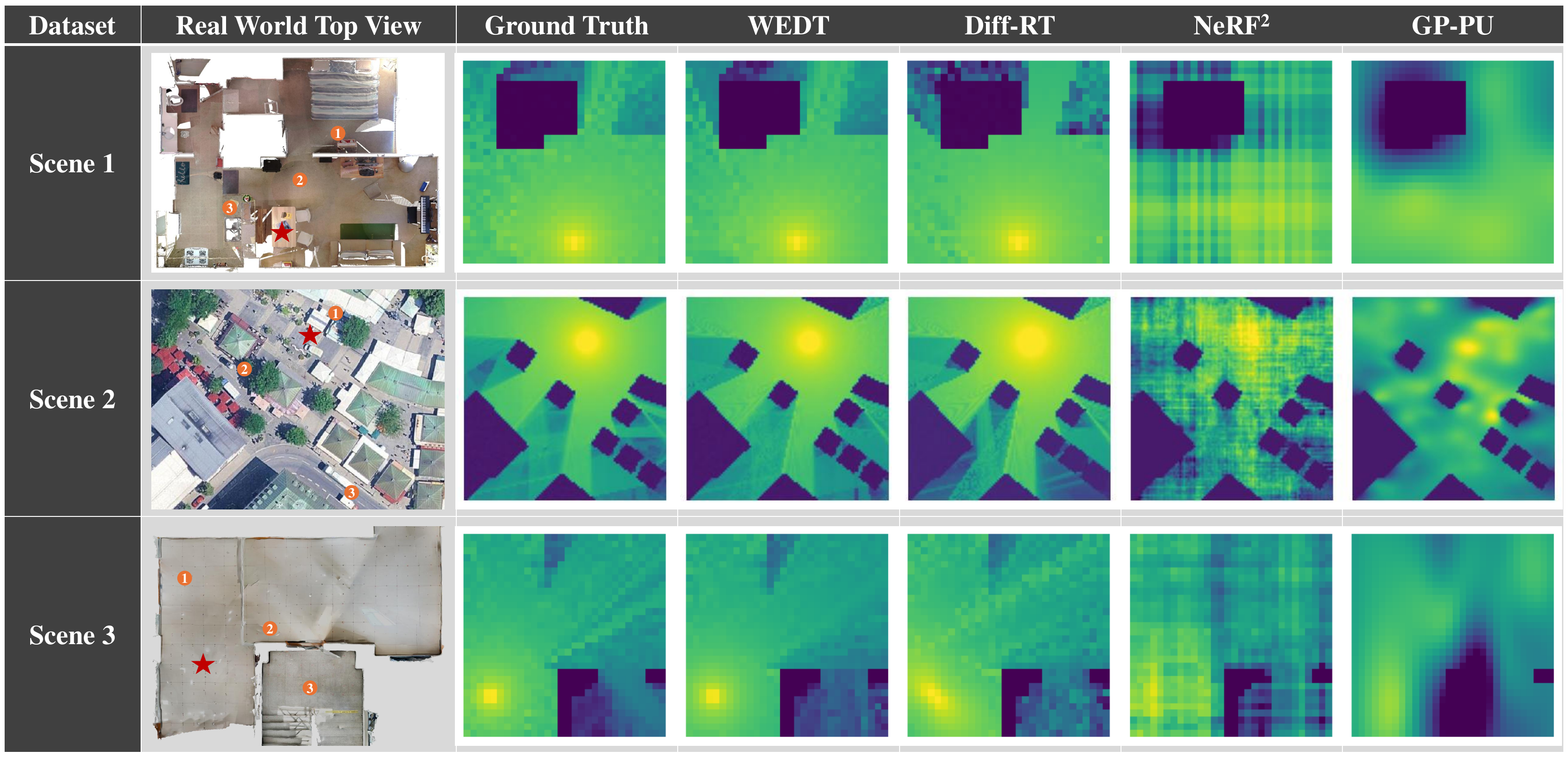}
    \caption{\textbf{Visualization of channel gain across different datasets for WEDT and various baselines.} Scene 1 and Scene 2 correspond to the indoor apartment and outdoor Munich scenes in the public dataset, respectively, while Scene 3 is a real-world corridor scene from Wireless Valley. The fixed TX location is marked with a red asterisk, and sparse training samples (M=30) are randomly distributed across each scene. The channel gains predicted by different methods are compared with the Ground Truth on the right.} 
    \label{fig_effective}
    \vspace{-10pt}
\end{figure*}

\begin{figure*}[t] 
    \centering
    \includegraphics[width=\textwidth]{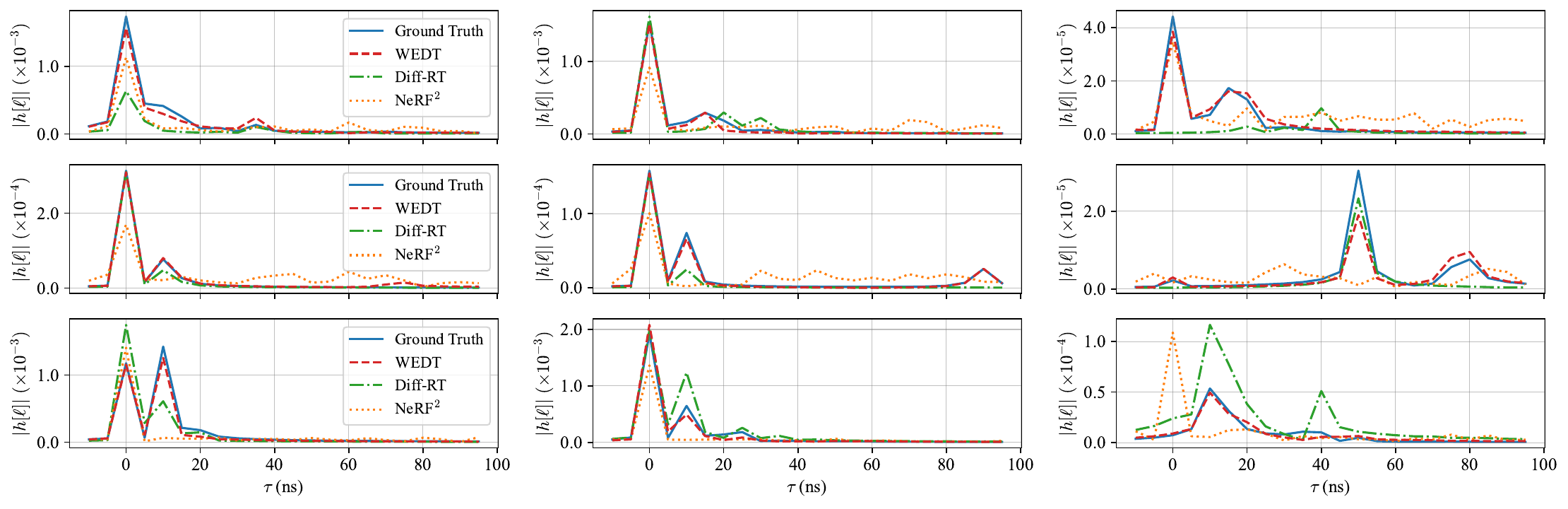}
    \caption{\textbf{Visualization of CIR across different datasets for WEDT and various baselines.} The three rows sequentially correspond to the three scenes. The subfigures from left to right in each row display the predicted CIRs at the locations marked by the numbered orange dots (positions 1, 2, and 3) shown in the top views of Fig. 5.} 
    \label{fig_effective_csi}
    \vspace{-10pt}
\end{figure*}

We first evaluate the performance of WEDT in the channel prediction task under the fixed-Tx topology, where the Tx is fixed at the red star in Fig. \ref{fig_effective}.
Under extremely sparse sampling conditions, the channel gain maps predicted by different models are shown in Fig. \ref{fig_effective}. 
Visually, the prediction of WEDT shows the closest visual agreement with the ground truth in both spatial fading trend and shadowing details.
Quantitative analysis indicates that the MALE of WEDT in each scene is as low as 1.56, 1.77, and 2.43, and the SSIM reaches up to 0.95, 0.92, and 0.89, respectively.
Meanwhile, over the entire prediction region, the MCS between the CSI predicted by WEDT and the Ground Truth reaches 0.88, 0.78, and 0.79, respectively, demonstrating that WEDT also maintains high consistency in the more fundamental CSI prediction.
Fig. \ref{fig_effective_csi} further illustrates this from the perspective of multipath channel characterization, where we visualize the CIRs at typical prediction points marked by orange dots in Fig. \ref{fig_effective} across the three scenes.
The results show that, compared with other baselines, WEDT can more accurately capture the arrival delays and relative energy amplitudes of the dominant multipath components.

In the above experiments, the channel prediction process with a fixed-Tx topology can be viewed as a typical base station to device (B2D) communication topology. 
Under the 6G Internet of Everything vision, direct device to device (D2D) communication is also an important topology that cannot be ignored.
Thus, we next examine whether the learned environment representation generalizes to unseen D2D transceiver links.
To this end, we directly transfer the models trained under the B2D topology to the D2D topology to evaluate their generalization performance.
Specifically, for the first two scenes, we randomly generate 100 new Tx positions and predict the channels over the entire scene. For Scene 3, limited by real-world measurement conditions, we collect 132 unseen and spatially randomly distributed Tx-Rx links, therefore the SSIM metric for the full coverage map is no longer applicable in this scene.
Table 1 details the generalization performance comparison among the models. 
The results show that, although the prediction performance of WEDT degrades to some extent on the unseen D2D topology relative to the B2D task, it consistently demonstrates stronger generalization capability than the baselines across all scenes.

\begin{table}[htbp]
\centering
\caption{Performance comparison between WEDT and the Baselines in D2D communications.}
\label{tab:performance}
\vspace{0.5em} 
\begin{tabular}{llccc}
\toprule
\textbf{Dataset} & \textbf{Method} & \textbf{MALE} $\downarrow$ & \textbf{SSIM} $\uparrow$ & \textbf{MCS} $\uparrow$ \\ \midrule

\multirow{4}{*}{\textbf{Scene 1}} 
    & WEDT  & 2.11 & 0.86 & 0.82 \\
    & Diff-RT   & 5.27 & 0.74 & 0.71 \\
    & NeRF$^2$  & 8.20 & 0.55 & 0.58 \\ \midrule

\multirow{4}{*}{\textbf{Scene 2}} 
    & WEDT  & 2.31 & 0.85 & 0.73 \\
    & Diff-RT       & 6.79  & 0.72 & 0.65 \\
    & NeRF$^2$       & 9.14 & 0.53 & 0.54 \\ \midrule

\multirow{4}{*}{\textbf{Scene 3}} 
    & WEDT  & 3.03  & -- & 0.75 \\
    & Diff-RT       & 5.84  & -- & 0.68 \\
    & NeRF$^2$       & 8.76  & -- & 0.55 \\
\bottomrule
\end{tabular}
\vspace{-10pt}
\end{table}

The performance advantage of WEDT comes from its environment-level representation. 
Rather than fitting link-specific channel responses, WEDT calibrates a scene-level EM property field within an explicit RT-based propagation framework using sparse real measurements and geometric priors. 
Because this calibrated field represents scene properties instead of a specific transceiver configuration, the resulting WEDT can be directly applied to unseen D2D links. 
In contrast, although Diff-RT also employs RT, direct parameter optimization under extremely sparse supervision is more prone to overfitting. 
NeRF$^2$ does not leverage geometric priors of the environment and struggles to stably learn effective implicit field representations under sparse supervision. 
GP-PU is essentially still a statistical interpolation method based on spatial correlation, which lacks modeling of the underlying propagation mechanism and cannot support high-dimensional complex CSI prediction or topology generalization tasks.
Additionally, we test the conventional ITU labeling model in the real-world scene. 
For the B2D prediction task, its corresponding metrics are only 4.64, 0.79, and 0.65, while for the D2D generalization task, the MALE and MCS values are 6.01 and 0.62, respectively. 
These results further show that, relative to idealized standard material parameters, the calibrated EM property field in WEDT provides a more realistic representation of the propagation environment.

\begin{figure}[!t]
  \centering
  \includegraphics[width=\linewidth]{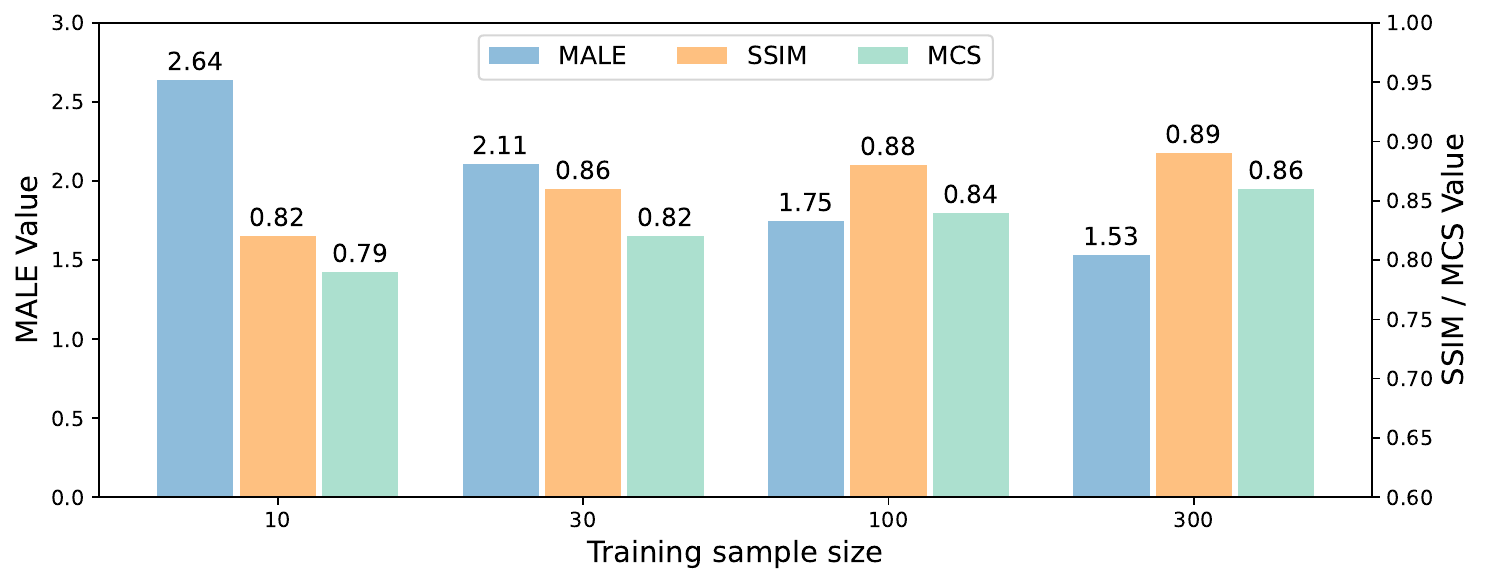}
  \caption{\textbf{Impact of the training sample size on the D2D generalization performance of WEDT in Scene 1.} The number of B2D training samples is set to 10, 30, 100, and 300. The left y-axis corresponds to MALE, while the right y-axis corresponds to SSIM and MCS.}
  \label{fig_fewshot}
\end{figure}

To further evaluate the sample efficiency of WEDT and its robustness under extremely sparse measurement conditions, we vary the B2D training sample size in Scene 1 and evaluate the performance of the resulting WEDT on the D2D generalization task.
The training sample size M is set to 10, 30, 100, and 300, with the corresponding results shown in Fig. \ref{fig_fewshot}.
Even with only 10 training samples available, WEDT still maintains reasonable prediction accuracy, indicating that the BCM provides useful probabilistic supervision for calibrating the EM property field in the low-data regime. 
As the training dataset size increases, the prediction performance of WEDT exhibits a steady positive gain, with MALE decreasing monotonically while SSIM and MCS steadily increase. 
This also reveals the self-learning potential of WEDT, as it can be continuously refined over time by collecting additional user measurements, thereby approaching the wireless environment more closely.

\begin{figure}[!t]
  \centering
  \includegraphics[width=\linewidth]{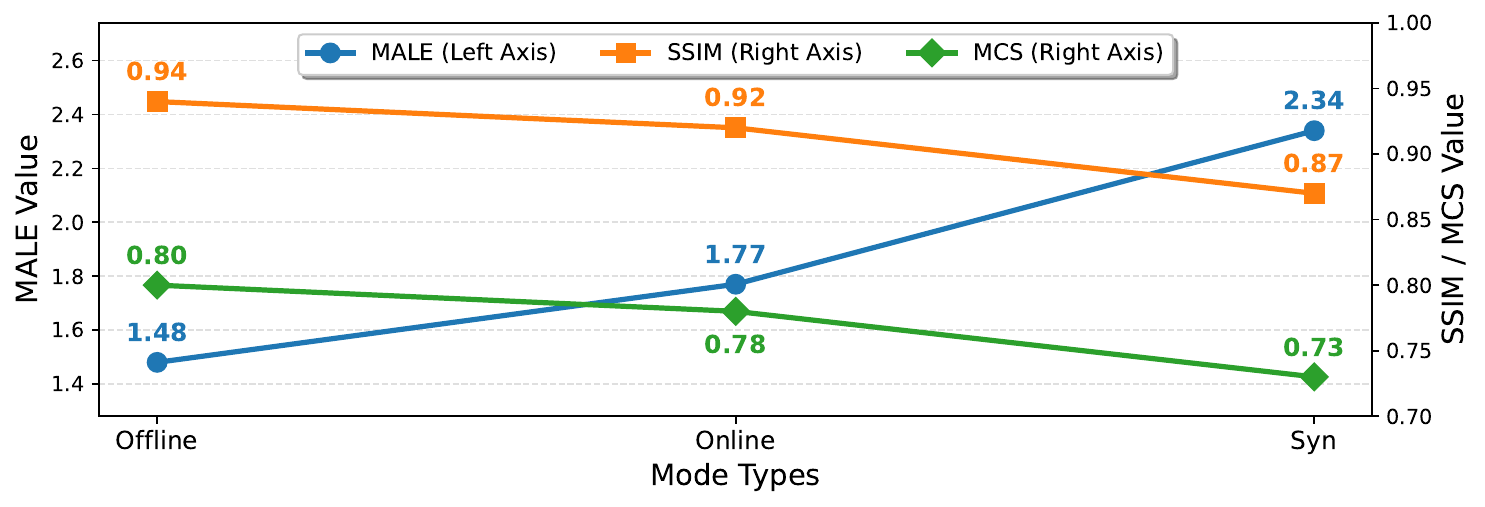}
  \caption{\textbf{Evaluation of the channel prediction performance of WEDT in Scene 2 under different data sampling modes.} The three sampling modes are the offline, online, and synchronous mode.}
  \label{fig_mode}
  \vspace{-10pt}
\end{figure}

The previous experiments use randomly sampled training data, which can be viewed as an online sampling mode based on crowdsourced data collected by active users in the network. 
However, this mode is often constrained by data collection cost, localization accuracy, and privacy protection. 
To evaluate the robustness of WEDT under different acquisition strategies, we further compare two other practical sampling modes in Scene 2: the offline mode and the synchronous mode. 
Specifically, in the former, engineers actively deploy transceivers to perform targeted channel sampling, while in the latter, channel data are collected simultaneously during 3D scene scanning and reconstruction. 
Fig. \ref{fig_mode} presents the channel prediction results under different sampling modes. 
It can be observed that the model performance exhibits a clear tiered pattern, decreasing in the order of the offline, online, and synchronous mode.
The offline mode can rely on expert knowledge to collect more representative measurement samples at critical locations in the environment, thus achieving the best performance. 
In contrast, the synchronous mode is constrained by the single trajectory of the scanning device, and the sampling points show stronger spatial correlation, which introduces more redundant information and leads to performance degradation, but it still remains highly competitive overall.
The above results show that WEDT can adapt to different data sampling modes and maintain stable prediction performance while preserving deployment flexibility.


\subsection{Application Evaluation}

\begin{figure}[!t]
  \centering
  \includegraphics[width=\linewidth]{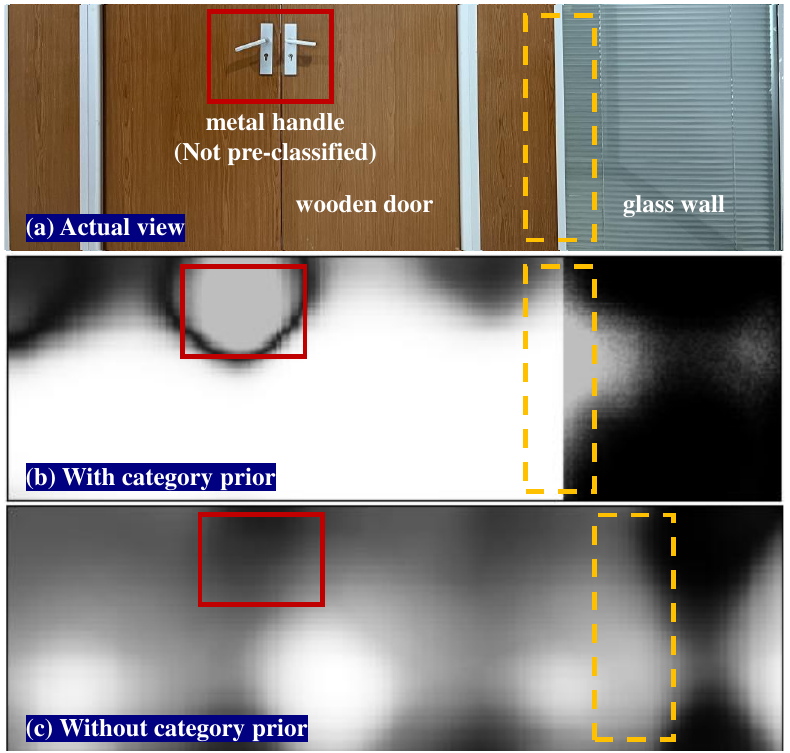}
  \caption{\textbf{Visualization of material-related spatial variations captured by the calibrated EM property field in Scene 3.} (a) Real view of the selected wall. The red box indicates the metal door handle, and the yellow box indicates the transition region between the wooden door and the glass wall. (b) Reflection coefficient of the EM property field with object category priors. (c) Reflection coefficient of the EM property field without object category priors.}
  \label{fig_vision}
  \vspace{-10pt}
\end{figure}

We next evaluate whether the calibrated EM property field captures material-related spatial variations in the environment, which is relevant to non-contact environment sensing in ISAC.  
To verify whether the calibrated EM property field can characterize material related spatial variations in the environment, we select a wall in Scene 3 for analysis, with its actual view shown in Fig. \ref{fig_vision}(a).
We trained two parameterized field representation models, one model uses both positional encoding and pre-classified object information, where the object categories include wooden doors and glass walls, and the other model uses only positional encoding.
We then query the calibrated EM property field on the spatial grid of this region, calculate the reflection coefficient by the Fresnel equation, and visualize the corresponding distributions, with the results shown in Fig. 9(b) and (c), respectively.

It can be observed that, after introducing object category priors, calibrated EM property field can clearly distinguish the macroscopic boundary between the wooden door and the glass wall. 
More importantly, although the metal door handle in the red boxed region is not further classified in the visual annotation and is still included in the wooden door region, EM property field still exhibits a distinct local anomalous response in its vicinity.
This indicates that the calibrated EM property field can not only characterize coarse-grained EM differences across different semantic regions, but also capture finer local variations within the same semantic region.
Meanwhile, even without object category priors, the calibrated EM property field also exhibits certain material-aware capabilities.
Specifically, a discernible transition region marked by the yellow box can still be observed between the wooden door and the glass wall, and some local contrast variation also remains around the door handle.
Although its result is smoother and the boundary is less clear, this phenomenon indicates that even without relying on semantic class labels, WEDT can still learn effective material related spatial structures solely from channel measurements.
The above results indicate that WEDT shows potential for non-contact environmental sensing in ISAC and shows promise for supporting applications in material sensing and local structural anomaly detection.

To further verify the practical application value of WEDT in the physical layer planning of 6G wireless systems, we conducted a WEDT-based RIS-assisted coverage enhancement experiment.
As shown in Fig. 10, we deployed a RIS operating at the 6.6 GHz band in scene 3 to enhance signal coverage for the adjacent room experiencing severe blockage from the Tx.
This RIS measures 32×32 cm and consists of 16×16 elements supporting 2-bit phase control, and more details on its parameters can be found in \cite{YX_RIS}.
We set eight candidate deployment positions spaced 0.2 m apart along the corridor wall.
While maintaining LoS paths with both the Tx and the target area, the RIS at each position employs a beam focusing coding strategy to steer the received Tx signal toward the center of the target area.

\begin{figure}[!t]
  \centering
  \includegraphics[width=\linewidth]{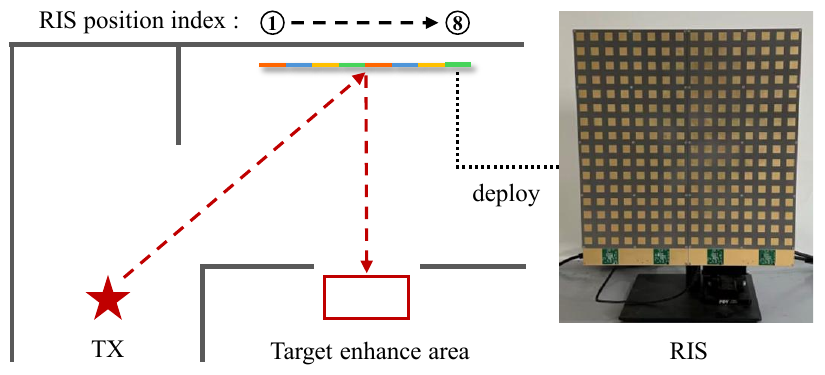}
  \caption{\textbf{RIS deployment experiment for coverage enhancement in Scene 3.} The eight candidate RIS positions are arranged from left to right with an interval of 0.2 m as shown in the figure, and all of them can ensure the existence of a LoS path between the TX and the target enhancement area. The right side shows the photograph of the RIS used in the experiment.}
  \label{fig_ris}
\end{figure}
\begin{figure}[!t]
  \centering
  \includegraphics[width=\linewidth]{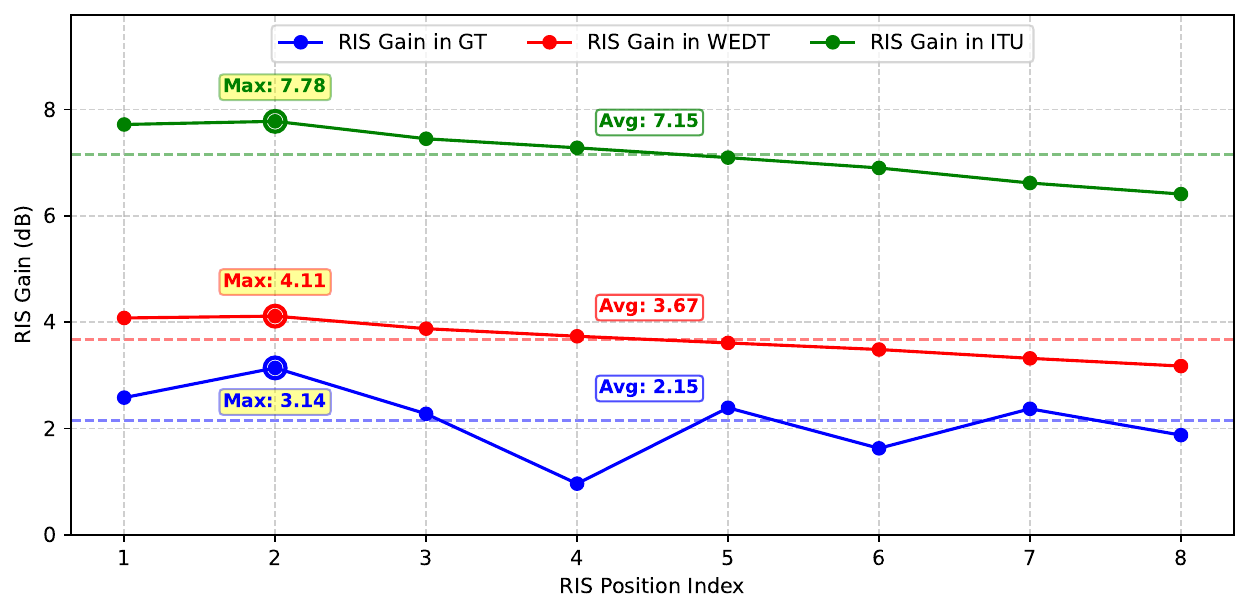}
  \caption{\textbf{Coverage enhancement gain after RIS deployment under measurements, WEDT simulation, and ITU labeling simulation.} The solid lines are the average gain over the target area at each RIS position. The dashed lines are the average gain across all RIS positions for each method.}
  \label{fig_ris_result}
  \vspace{-10pt}
\end{figure}

Concurrent with the on-site measurements in scene 3, we conducted simulation evaluations of the RIS deployment in WEDT and ITU labeling model.
Fig. 11 shows the average channel gain obtained in the target area under different methods after deploying the RIS.
The results indicate that both WEDT and the ITU labeling model correctly perform spatial optimization and accurately predict position 2 as the optimal deployment point to maximize coverage gain, which is consistent with the empirical measurements.
However, the two models exhibited significant differences when evaluating the absolute gain.
In measurements, the average RIS deployment gain across eight locations was 2.15 dB, whereas the ITU-assigned model yielded an average gain as high as 7.15 dB, indicating severe overestimation, while the 3.67 dB predicted by WEDT is numerically closer to the actual physical environment.
This indicates that the closed-loop calibrated WEDT can more accurately reflect the impact of real environmental EM properties on RIS deployment performance.
These results indicate that WEDT can correctly rank candidate RIS locations and provides a much more realistic estimate of the achievable coverage gain than the ITU-based model. Therefore, RIS simulation in WEDT can support more credible pre-deployment evaluation and help network operators better balance expected performance improvement against deployment cost.

Finally, we evaluate the capability of WEDT as a high-fidelity synthetic data engine for supporting downstream AI models. 
Taking the recently proposed deep generative model RadioFlow \cite{Radioflow} as an example, this model is built on the principle of flow-based generation and can achieve millisecond-level end-to-end channel gain map inference on edge devices.
Specifically, we use WEDT and the ITU labeling model to generate large-scale synthetic channel gain map datasets for random transmitter positions, and train two RadioFlow models on these two types of datasets, respectively. 
We then test both models on the real dataset of Scene 3 in the above D2D experiment to evaluate how different synthetic datasets affect their generalization performance in real environments.

\begin{figure}[!t]
  \centering
  \includegraphics[width=\linewidth]{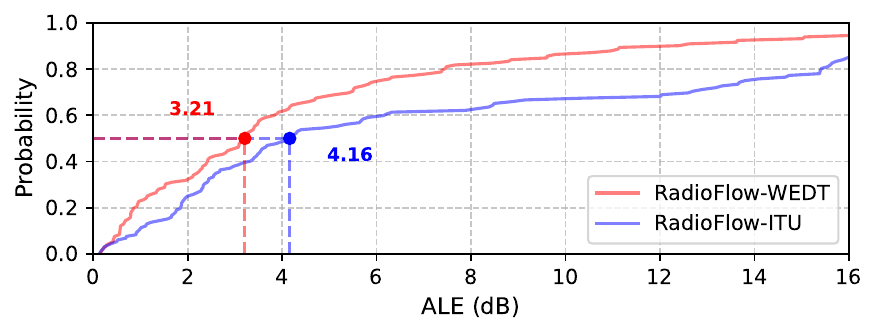}
  \caption{\textbf{CDF of the ALE for channel gain prediction achieved by RadioFlow trained on synthetic data generated based on WEDT and ITU labeling model.} The test set is the D2D measurement dataset in Scene 3 and the marked points indicate the median ALE.}
  \label{fig_radioflow}
  \vspace{-10pt}
\end{figure}

Fig. 12 shows the cumulative distribution function (CDF) of the absolute logarithmic error (ALE) for RadioFlow trained on the two datasets in the channel gain prediction task.
The results show that RadioFlow-WEDT trained on WEDT synthetic data outperforms RadioFlow-ITU trained on ITU synthetic data over most error ranges, and its ALE CDF curve lies further toward the upper left overall, indicating smaller prediction errors and higher reliability in real environment testing.
The numerical results show that the median ALE of RadioFlow-WEDT is 3.21 dB, whereas that of RadioFlow-ITU increases to 4.16 dB.
This clear performance gap indicates that the traditional ITU labeling method has a clear mismatch with the real EM environment, and therefore AI models trained on synthetic data generated by this model inherit such modeling bias, which limits their generalization capability.
In contrast, WEDT learns a calibrated EM property field through closed-loop calibration, which yields synthetic data that are more consistent with the real propagation environment.
This reduces the sim-to-real mismatch seen by the downstream model and improves synthetic-to-real transfer.

\section{CONCLUSION}

In this paper, we studied WEDT construction under sparse measurements. By calibrating a scene-level EM property field on top of a reconstructed geometric DT, the proposed construction paradigm evolves a geometry-consistent scene representation into a wireless environment representation that is more consistent with real propagation observations. Instead of relying on direct link fitting, it uses geometry-prior probabilistic supervision together with differentiable RT-based channel computation to make such calibration feasible under limited CSI. Results in both public and real-world scenes show that WEDT achieves accurate channel prediction, transfers well to unseen transceiver topologies, and remains robust across different sampling conditions. The resulting WEDT also demonstrates value in material-related environment sensing, physical-layer planning, and synthetic data generation for downstream wireless AI tasks. These findings confirm the value of the proposed paradigm for propagation-consistent WEDT construction and related wireless applications. Future work could explore dynamic environments adaptation and online updating to fully exploit the potential of WEDT.


\bibliographystyle{IEEEtran}
\bibliography{refs}

\vfill

\end{document}